\documentclass[preprint,aps,prb,floatfix,a4paper,showpacs]{revtex4}
\usepackage{amsfonts}
\usepackage{amssymb}
\usepackage{graphicx}

\input{tcilatex}

\begin{document}

\title{Enhancement of nonclassical properties of two qubits via deformed
operators}
\author{N. Metwally$^{1,2}$, M. Sebawe Abdalla$^{3}$ and
M. Abdel-Aty$^{1,4}$}
\affiliation{$^{1}${\small Mathematics
Department, Faculty of
Science, South Valley University, Asswan, Egypt }\\
$^{2}${\small Mathematics Department, College of Science, Bahrain
University, 32038 Kingdom of Bahrain}
\\
$^{3}${\small \ Department of Mathematics, College of Science, King
Saud University, PO. Box 2455, Riyadh 11451, Saudi Arabia}
\\
$^{4}${\small Mathematics Department, Faculty of Science, Sohag
University, 82524 Sohag, Egypt}
 }
\date{\today}

\begin{abstract}
We explore the dynamics of two atoms interacting with a cavity field
via deformed operators. Properties of the asymptotic regularization
of entanglement measures proving, for example, purity cost,
regularized fidelity and accuracy of information transfer are
analyzed. We show that the robustness of a bipartite system having a
finite number of quantum states vanishes at finite photon numbers,
for arbitrary interactions between its constituents and with cavity
field. Finally it is shown that the stability of the purity and the
fidelity is improved in the absence of the deformation parameters.
\end{abstract}

 \maketitle

\section{Introduction}

In the recent past, there has been a great deal of interest in the
study of
information transfer in connection between several physical fields \cite%
{nie00}. Using classical communication between two spatially
separated parties, Pramanik et al \cite{pra10} have introduced a new
scheme which is implemented by a sequence of unitary operations
along with suitable spin-measurements. This protocol demonstrated
the possibility of using intra-particle entanglement as a physical
resource for performing information theoretic tasks. Di Franco and
others \cite{fra10} have proposed a strategy for perfect state
transfer in spin chains based on the use of an un-modulated coupling
Hamiltonian whose coefficients are explicitly time dependent. They
have shown that, if specific and non-demanding conditions are
satisfied by the temporal behavior of the coupling strengths, their
model allows perfect state transfer.

In a one-dimensional coupled resonator waveguide, an efficient
scheme for
the implementation of quantum information transfer has been given \cite{li09}%
. It has shown that quantum information could be transferred
directly between the opposite ends of the coupled waveguide without
involving the intermediate nodes via either Raman transitions or the
stimulated Raman adiabatic passages. Further, quantum information
transfer from spin to orbital angular momentum of photons has been
also discussed by Nagali et al \cite{nag09}. On the other hand, many
physical applications have been
investigated on the basis of the $q$-deformation of the Heisenberg algebra (%
\cite{lav08}-\cite{fin96}). In Ref. \cite{lav06}, a $q$-deformed
Poisson bracket, invariant under the action of the $q$-symplectic
group, has been derived and a classical $q$-deformed
thermostatistics has been proposed in Ref. \cite{lav07}. In this
sense one can say that, quantum entanglement is fundamental in
quantum physics both for its essential role in understanding the non
locality of quantum mechanics \cite{ein35,bel64} as well as for its
practical application in quantum information processing
\cite{nie00,ben00}. It is a kind of counter intuitive non local
correlation.

The main purpose of this paper is to examine the effect of the $q$%
-deformation on transferring the information between two parties.
This may be achieved if we considered one of the most generalized
Hamiltonian which describes the interaction between pair of qubits
and an electromagnetic
field. In this case it is essential to include the effect of the $q$%
-deformation in the structure of the Hamiltonian through the
dynamical operators. In the meantime, one of our target is to see
the effect of the multiphoton process on the system. This means
that, the model we plane to
introduce should contains the effect of multiphoton in addition to the $q$%
-deformation. For this reason we devote the next section to
introduce our model and to give the solution for the time-dependent
wave function. In section \textbf{III} we consider the purity from
which we discuss the degree of mixture for the system. We devote
section \textbf{IV} to examine the accuracy of the information
transfer against the scaled time. This is followed by our
consideration of the population in section \textbf{V}. Finally we
give our conclusion in section \textbf{VI}.

\section{The model}

We consider the interaction between two qubits and a multiphoton
cavity field. This Hamiltonian can be written in the following form
\begin{equation}
\frac{\mathrm{\hat{H}}}{\hslash }=\omega \hat{A}^{\dagger }\hat{A}%
+\sum_{i=1}^{2}\left( \frac{\Omega _{i}}{2}\hat{S}_{z}^{(i)}+\lambda
\left( \hat{S}_{+}^{(i)}\hat{A}^{m}+\hat{S}^{(j)}\hat{A}^{\dagger
m}\right) \right) ,  \label{1}
\end{equation}%
where $\omega $ is the frequency of the field. $\hat{A}$ and $\hat{A}%
^{\dagger }$ are the deformed annihilation and creation operators
defined by
\begin{equation}
\hat{A}=\hat{a}f(\hat{n}),\quad \ \hat{A}^{\dagger }=f(\hat{n})\hat{a}%
^{\dagger },  \label{2}
\end{equation}%
where $f(\hat{n})$ is a function of the number operator $\hat{n}=\hat{a}%
^{\dagger }\hat{a}.$ The operators $\hat{a}$ and $\hat{a}^{\dagger
}$ are
the usual bosonic annihilation and creation operators with the property $%
\left[ \hat{a},\hat{a}^{\dagger }\right] \ =1$ In the meantime the
deformed operators $\hat{A}$ and $\hat{A}^{\dagger }$ satisfy the
commutation relation
\begin{eqnarray}
\left[ \hat{A},\hat{A}^{\dagger }\right] &=&\left( \hat{n}+1\right)
f^{2}\left( \hat{n}+1\right) -\hat{n}f^{2}\left( \hat{n}\right)
\text{,}
\nonumber \\
\left[ \hat{A},\hat{n}\right] &=&\hat{A},\qquad \left[ \hat{A}^{\dagger },%
\hat{n}\right] =-\hat{A}^{\dagger }.  \label{3}
\end{eqnarray}%
The operators $\hat{S}_{\pm }^{(i)}$ and $\hat{S}_{z}^{(i)},i=1,2$
are the usual raising (lowering) and inversion operators for the
two-level atomic system, satisfying the relations
\begin{equation}
\left[ \hat{S}_{z}^{(i)},\hat{S}_{\pm }^{(j)}\right] =\pm
2\hat{S}_{\pm
}^{(i)}\delta _{ij}\qquad \text{and\qquad }\left[ \hat{S}_{+}^{(i)},\hat{S}%
_{-}^{(j)}\right] =\hat{S}_{z}^{(i)}\delta _{ij},  \label{4}
\end{equation}%
where $\delta _{ij}$ is the Kronecker delta so that $\delta _{ij}=1$
if $i=j$ and $zero$ otherwise. To reach our goal we have to find the
explicit expression of the wave function in the Schr\"{o}dinger
picture. For this reason we employ the Heisenberg equations of
motion together with the Hamiltonian (\ref{1}) that to derive some
constants of motion from which we are able to achieve our task. In
this case we have
\begin{eqnarray}
\frac{d{\hat{S}_{z}^{(i)}}}{dt} &=&2i\lambda (\hat{S}_{-}^{(i)}\hat{A}%
^{\dagger m}-\hat{S}_{+}^{(i)}\hat{A}^{m}),\qquad \frac{d\hat{A}}{dt}%
=-i\omega \hat{A}-i\lambda m\hat{A}^{\dagger
(m-1)}\sum_{i=1}^{2}S_{-}^{(i)},
\nonumber \\
\frac{d\hat{A}^{\dagger }}{dt} &=&i\omega \hat{A}^{\dagger }+i\lambda m\hat{A%
}^{(m-1)}\sum_{i=1}^{2}S_{-}^{(i)}.  \label{5}
\end{eqnarray}%
From the above equations we can deduce that
\begin{equation}
\hat{M}=\hat{N}+\frac{m}{2}\sum_{i=1}^{2}{\hat{S}_{z}^{(i)},\qquad }\hat{N}=%
\hat{A}^{\dagger }\hat{A},  \label{6}
\end{equation}%
where $\hat{M}$ is a constant of motion and $\hat{N}$ is the number
of photon of the deformed operator $\hat{A}$. Using Eq.(\ref{6}),
the deformed Hamiltonian $\mathrm{\hat{H}}$ becomes,
\begin{equation}
\frac{\mathrm{\hat{H}}}{\hslash
}=\hat{M}+\sum_{j=1}^{2}\hat{C}_{j},\quad
\text{where\qquad }\hat{C}_{j}=\frac{\Delta _{j}}{2}\hat{S}%
_{z}^{(i)}+\lambda \left( \hat{S}_{+}^{(i)}\hat{A}^{m}+\hat{S}^{(j)}\hat{A}%
^{\dagger m}\right) ,  \label{7}
\end{equation}%
where $\Delta =\Omega _{j}-m\omega $ is the detuning parameter.
Since the Hamiltonian given by equation (\ref{1}) is a constant of
motion, therefore
the operator $\hat{C}_{j}$ is also constant of motion and consequently $\hat{%
M}$ and $\hat{C}_{j}$ are commute. To find the state vector $|\Psi
(t)\rangle $ we assume that the initial atomic state takes the form
\begin{equation}
|\Psi (0)\rangle _{1,2}=a_{1}|e,e\rangle +a_{2}|e,g\rangle
+a_{3}|g,e\rangle +a_{4}|g,g\rangle ,  \label{8}
\end{equation}%
where the suffix $1$ and $2$ refers to the first and the second
qubit, while $|e\rangle $ and $|g\rangle $ are the excited and
ground states, respectively. It should be noted that
$a_{i},i=1,2,3,4$ are arbitrary complex quantities that satisfy the
condition$\sum_{i=1}^{4}|a_{i}|^{2}=1.$ Now suppose we consider the
field is prepared in the coherent state
\begin{equation}
|\alpha \rangle =\sum_{n=0}^{\infty }Q_{n}|n\rangle \qquad \text{and\qquad }%
Q_{n}=\frac{\alpha ^{n}}{\sqrt{n!}}\exp \left( -\frac{1}{2}|\alpha
|^{2}\right) .  \label{9}
\end{equation}%
Therefore, the initial state of the qubits and the field takes the
form
\begin{equation}
|\Psi (0)\rangle _{s}=\sum_{n=0}^{\infty }Q_{n}\left( |\Psi
(0)\rangle _{1,2}\right) \otimes |n\rangle .  \label{10}
\end{equation}%
In this case if we use the Schr\"{o}dinger equation $i\hslash
\partial |\Psi
\rangle /\partial t=\mathrm{\hat{H}}|\Psi \rangle $ and the Hamiltonian (\ref%
{7}) together with the above equation, then after some calculations
the state vector of the system at $t>0$ can be written thus
\begin{equation}
|\Psi (t)\rangle _{s}=\sum_{n=0}^{\infty }A_{n}(t)|e,e,n\rangle
+B_{n}(t)\left( |e,g,n+2\rangle +C_{n}(t)|g,e,n+2\rangle \right)
+D_{n}(t)|g,g,n+4\rangle ,  \label{11}
\end{equation}%
where $A_{n}(t),B_{n}(t),C_{n}(t)$ and $D_{n}(t)$ are complex
time-dependent functions have the expressions
\begin{eqnarray}
A_{n}(t) &=&a_{1}Q_{n}-\nu _{1}\left( a_{1}\nu _{1}Q_{n}+a_{4}\nu
_{2}Q_{n+2}\right) \frac{\sin ^{2}\mu _{n}t}{\mu _{n}^{2}}-i\nu
_{1}(a_{2}+a_{3})Q_{n+1}\frac{\sin 2\mu _{n}t}{2\mu _{n}t},  \nonumber \\
B_{n}(t) &=&Q_{n+1}\left( a_{2}\cos ^{2}\mu _{n}t-a_{3}\sin ^{2}\mu
_{n}t\right) -i\left( a_{1}\nu _{1}Q_{n}+a_{4}\nu _{2}Q_{n+2}\right) \frac{%
\sin \mu _{n}t}{2\mu _{n}t},  \nonumber \\
C_{n}(t) &=&Q_{n+1}\left( a_{3}\cos ^{2}\mu _{n}t-a_{2}\sin ^{2}\mu
_{n}t\right) -i\left( a_{1}\nu _{1}Q_{n}+a_{4}\nu _{2}Q_{n+2}\right) \frac{%
\sin \mu _{n}t}{2\mu _{n}t},  \nonumber \\
D_{n}(t) &=&a_{4}Q_{n+2}-\nu _{1}\left( a_{1}\nu _{2}Q_{n}+a_{4}\nu
_{2}Q_{n+2}\right) \frac{\sin ^{2}\mu (n)t}{\mu _{n}^{2}}-i\nu
_{2}(a_{2}+a_{3})Q_{n+1}\frac{\sin 2\mu _{n}t}{2\mu _{n}t}.  \nonumber \\
&&  \label{12}
\end{eqnarray}

In the above equation we have used the abbreviations
\begin{eqnarray}
\nu _{1}(n) &=&\lambda \sqrt{\frac{\left( n+m\right) !}{n!}G\left(
n+m\right) },\quad \nu _{2}(n)=\lambda \sqrt{\frac{\left( n+m\right) !}{n!}%
G\left( n+2m\right) },  \nonumber \\
\mu (n) &=&\frac{1}{\sqrt{2}}\sqrt{\nu _{1}^{2}(n)+\nu
_{2}^{2}(n)},\quad
G\left( n+m\right) =f(n)f(n+1)f(n+2).......f(n+m).  \nonumber \\
&&  \label{13}
\end{eqnarray}%
It should be noted that in the previous calculations we have
restricted ourself with the resonance case such that $\Delta =0.$
This is due to the fact that the solution in presence of the
detuning parameter ($\Delta \neq 0$ ) is too cumbersome to be
handled. Having obtained the state vector of the
system, we are therefore in a position to find the density matrix $\hat{\rho}%
=|\Psi \rangle \langle \Psi |.$ On the other hand, there are several
types
of the function $f(n)$ would causes a deformation, one of them is called $q$%
-deformation. In this context we restrict ourselves with a case in
which the function$\ f(n)$ is defined by
\begin{equation}
f(n)=\sqrt{\frac{1-q^{n}}{n(1-q)}}.  \label{14}
\end{equation}%
This in fact would help us to examine the effect of the
$q$-parameter on the present system during our discussion for some
statistical aspects.

\section{Degree of mixture}

It is well known that, if the degree of mixture increases, then the
quantum states tend to have a small amount of the entanglement. In
the meantime for the case of pair of qubits, the states with a large
enough degree of mixture are always separable \cite{kar,Mun,bat}.
There are several methods to measure the degree of mixture, among
these methods we mention here, the von Neumann entropy of the state
which is given by $\mathcal{S}=\hat{\rho}\ln \hat{\rho}$, this in
addition to the linear entropy \cite{Bose} besides the purity
$\mathcal{P}$. Since the tasks of quantum information and
computations require transferee information between two different
locations as in quantum teleportation \cite{Ben, bosc,Lee} or
between two nodes as in quantum computations \cite{kay}, therefore
we have to investigate the behavior of the purity of the carrier of
the information $\mathcal{P}$. This means that, the later method can
be adopted to discuss the degree of mixture for the present system,
where $\mathcal{P}=Tr\{\hat{\rho}^{2}\}$ and
$\hat{\rho}$ is the density operator. In this case if one uses equation (\ref%
{11}) the degree of purity for the qubits takes the form%
\begin{eqnarray}
\mathcal{P}(t) &=&|A_{n}(t)|^{2}+|B_{n}(t)|^{2}\left(
1+|A_{n+2}(t)|^{2}+2|C_{n}(t)|^{2}\right)  \nonumber \\
&&+|C_{n}|^{2}\left( 1+|A_{n+2}(t)|^{2}\right) +|D_{n}(t)|^{2}\left(
1+2|A_{n+4}(t)|^{2}+2|B_{n+2}(t)|^{2}\right) .  \label{15}
\end{eqnarray}%
In what follows we investigate the effect of the deformed operators
on the degree of the purity $\mathcal{P}$ provided that, the system
is initially prepared in one of the Bell's states. This type of
states is defined as a maximally entangled quantum state of two
qubits \cite{nie00}. In this context, we compare the effect of the
deformed operator on $\hat{\rho}_{\psi }=|\psi \rangle \langle \psi
|$ and $\hat{\rho}_{\phi }=$ $|\phi \rangle \langle \phi |$, where
\begin{equation}
|\psi \rangle =\frac{1}{\sqrt{2}}\left( |e,g\rangle +|g,e\rangle
\right) \qquad \text{and}\qquad |\phi \rangle
=\frac{1}{\sqrt{2}}\left( |e,e\rangle +|g,g\rangle \right) .
\label{16}
\end{equation}%
As one can see it is quite difficult to analyze equation (\ref{15}),
this is
due to the complication in the expression of the functions $%
A_{n}(t),B_{n}(t),C_{n}(t)$ and $D_{n}(t).$ For this reason we plot
some figures to display the behavior of the purity related to the
state $|\psi
\rangle $. For example in figures (1) we have plotted the purity $\mathcal{P}%
(t)$ against the scaled time $\lambda t$ for different values of the
involved parameters. In the absence of the effect of the
$q$-deformation,
namely $f(\hat{n})=1$ and for a fixed value of the mean photon number $%
|\alpha |^{2}=10,$ we have considered the cases in which the number
of
photons $m=1,2.$ In this case the purity starts with its maximum ( $\mathcal{%
P}(t)=1$ ) and as the time increases as its value decreases.
Moreover, we observe irregular rapid fluctuations, this is depicted
in Fig.(1a). Increasing the value of the number of photon $m=2,$ the
function shows also rapid fluctuations, however, with interference
between the pattern. In fact these fluctuations occur between the
maximum value of the purity and its minimum around the value $0.45,$
see Fig.(1b). On the other hand, the behavior of the function is
different when the $q$-deformation takes place, this is seen in
Figs.(1c,d). In this case we have considered different values for
the $q$ parameter, $q=0.1,0.5$ and $0.9.$ As we can see the general
behavior of the purity is the same for all the cases, where the
function decreases its value as the time increases, see Fig.(1c).

\begin{figure}[tbp]
\begin{center}
\includegraphics[width=15pc,height=12pc]{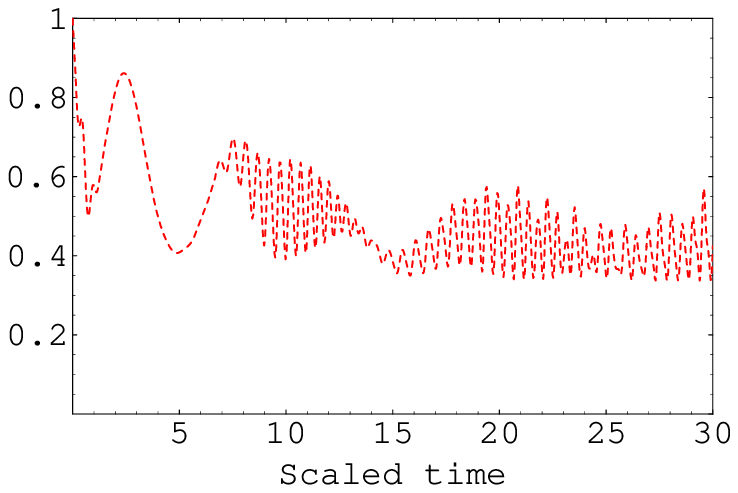} %
\includegraphics[width=15pc,height=12pc]{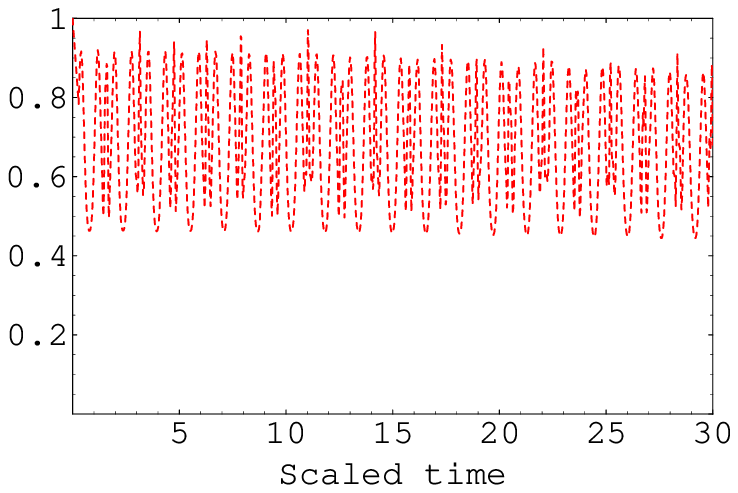} \put(-220,120){$%
(a)$}\put(-30,120){$(b)$} \put(-380,75){$\mathcal{P}$} \put(-180,75){$%
\mathcal{P}$}\\[0pt]
\includegraphics[width=15pc,height=12pc]{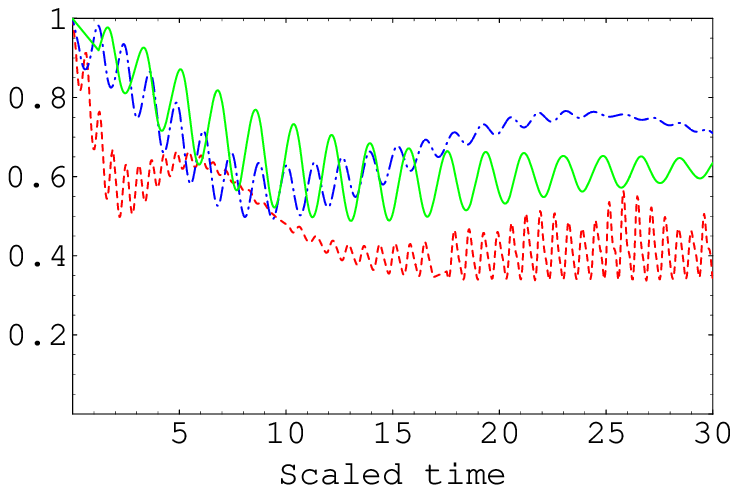} %
\includegraphics[width=15pc,height=12pc]{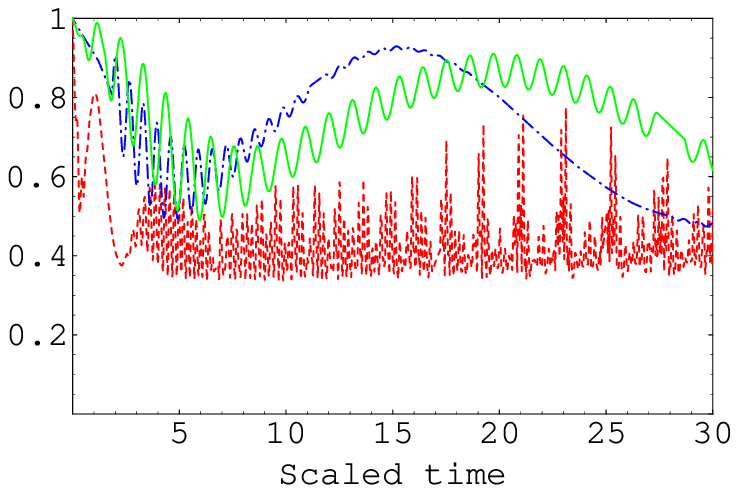}
\put(-220,120){$(c)$}\put(-30,120){$(d)$}\put(-380,75){$\mathcal{P}$}
\put(-180,75){$\mathcal{P}$}
\end{center}
\caption{The purity $\mathcal{P}$ for the state $|\protect\psi
\rangle $ where $\bar{n}=10$. (a) $m=1$ (b)$m=2$ (c)$m=1$, the
dash-dot, solid and the dot curves for $q=0.1,0.5,0.9$
respectively(d) the same as (c) but $m=2$.}
\end{figure}
In the meantime, we observe different behaviour for each case
individually. For small values of the deformity parameter $q$, the
amplitudes of the oscillations are small while the maximum values of
the purity are too large. As one increases the deformity parameter,
the purity function fluctuates rapidly and the amplitude of the
fluctuations are large. This behavior causes a decrease of the
minimum value of the purity as shown in Fig.(1c). More preciously
when we consider $q=0.1$ (dash-dot curve), the purity displays
regular fluctuations with an decrease in its value up to a certain
limit. Then the function rebounds to increase its value without
reaching its maximum. Similar behaviour is reported for the case in
which $q=0.5$ (solid curve ), however, the increment in its value is
less than that the previous case. More increases in the value of the
$q$-deformation parameter leads to more decrease in the value of the
purity. This is observed for the case in which $q=0.9$ (dot curve),
where we can also see irregular fluctuations around $0.4$ after
considerable reduction in its value. For $m=2,$ the
purity reduces its value after a short period of the time when we consider $%
q=0.1$ and $q=0.5$, however, after onset of the interaction for
$q=0.9$. As previously mentioned the function shows irregular
fluctuations with an decrease in its value as the time increases.
For the cases in which $q=0.1$ and $q=0.5$ the function turns to
increase its value after a period of the time shorter than that the
case in which $m=1 $. Also we observe that, for small values of $q$
the purity oscillates faster but with small amplitudes. This
behavior improves the purity, where its minimum value is always
greater than that the free deformations. Here we can report that for
$q=0.9$ there are more rapid fluctuations compared with the case in
which $m=1.$ The same behaviour can also reported for the other two
values for the $q$-deformation when $m=2$. In the meantime, the
minimum values of these fluctuations are nearly bounded as can be
seen in Fig.(1d). Now we turn our attention to
display the dynamics of the purity for the second type of Bell state vector $%
|\phi \rangle $. In this case and for $m=1$ we observe that, there
is a reduction in the value of the purity for all values of the
deformed parameter $q$, however, with a different ratio. For example
the function
decreases its value after onset of the interaction for the case in which $%
q=0.9,$ while for the other two cases $q=0.1$ and $q=0.5,$ the
reduction occurs after a short period of the time. Moreover, the
reduction for the case in which $q=0.9$ is too large compared with
the other two cases. In the meantime the rapid fluctations occurs in
this case too, see Fig.(2a). When we examine the case in which $m=2$
the purity shows more decreases in its value with rapid fluctuations
for all the values of the $q$ parameter. However, it is noted that
for the case in which $q=0.1$ there is a sudden change in the
function behaviour after considerable value of the time. In this
case the function decreases its value without fluctuations after
nearly half period of the considered time, see Fig.(2b).

From figures (1) and (2), we can conclude that, the decreasing of
the purity of the maximum entangled states $|\psi \rangle $ and
$|\phi \rangle $ is due to their interaction with the cavity mode.
Although an increase in the number of photons within cavity leads to
decrease the stability of the purity, however, it also improves its
maximum and minimum values. Therefore, the deformed operators would
enhance the purity of the traveling entangled states . Also we can
report that, when the number of photons increases then the deformity
parameter plays the role of error corrections with high efficiency.

\begin{figure}[tbp]
\begin{center}
\includegraphics[width=15pc,height=12pc]{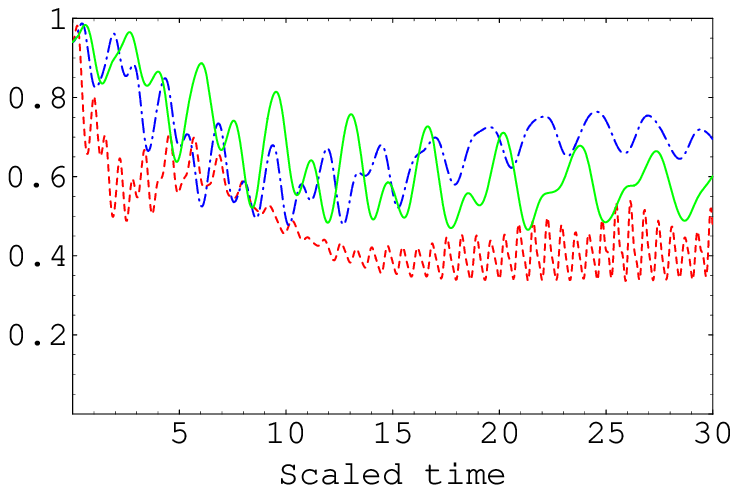} %
\includegraphics[width=15pc,height=12pc]{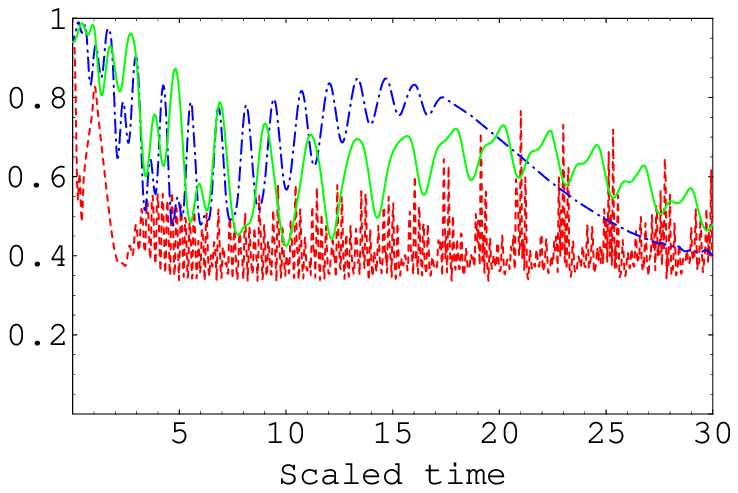}
\put(-220,120){$(a)$} \put(-30,120){$(b)$}
\put(-380,75){$\mathcal{F}$} \put(-180,75){$\mathcal{F}$}
\end{center}
\caption{The same as Fig.(1c) and (1d) respectively but for the
state vector $\bigl|\protect\phi \bigr\rangle$}
\end{figure}

\begin{figure}[tbp]
\begin{center}
\includegraphics[width=15pc,height=12pc]{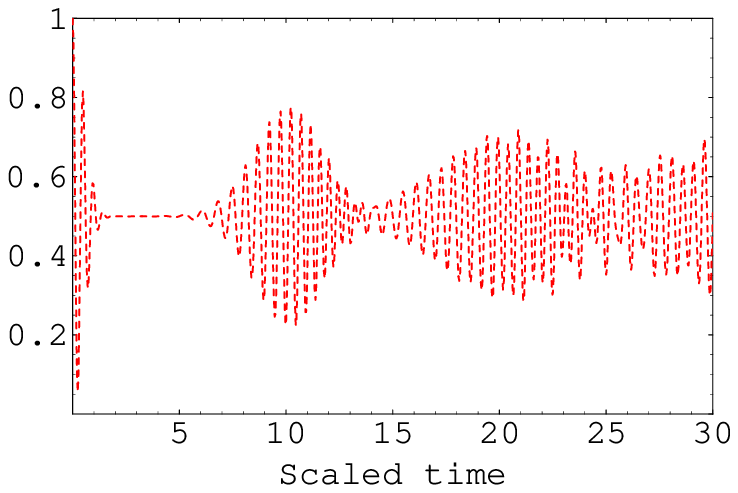} %
\includegraphics[width=15pc,height=12pc]{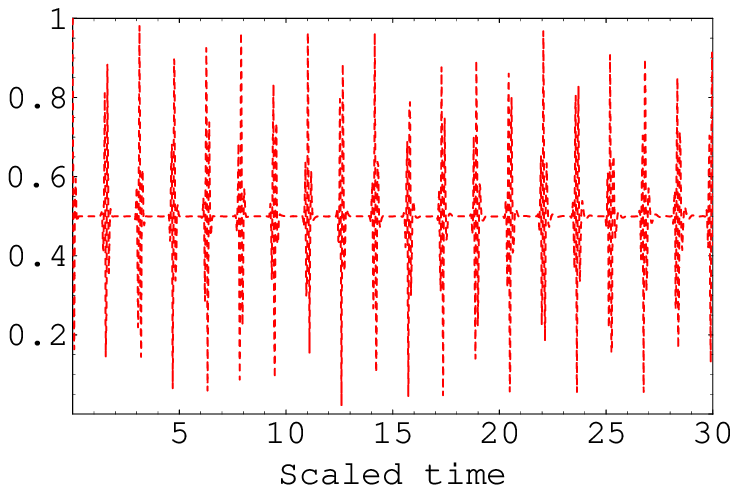} \put(-220,120){$%
(a)$}\put(-30,120){$(b)$} \put(-380,75){$\mathcal{F}$} \put(-180,75){$%
\mathcal{F}$}\\[0pt]
\includegraphics[width=15pc,height=12pc]{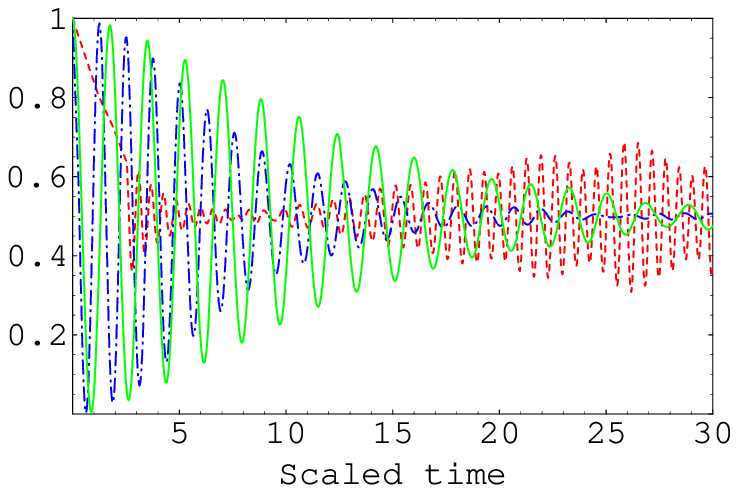} %
\includegraphics[width=15pc,height=12pc]{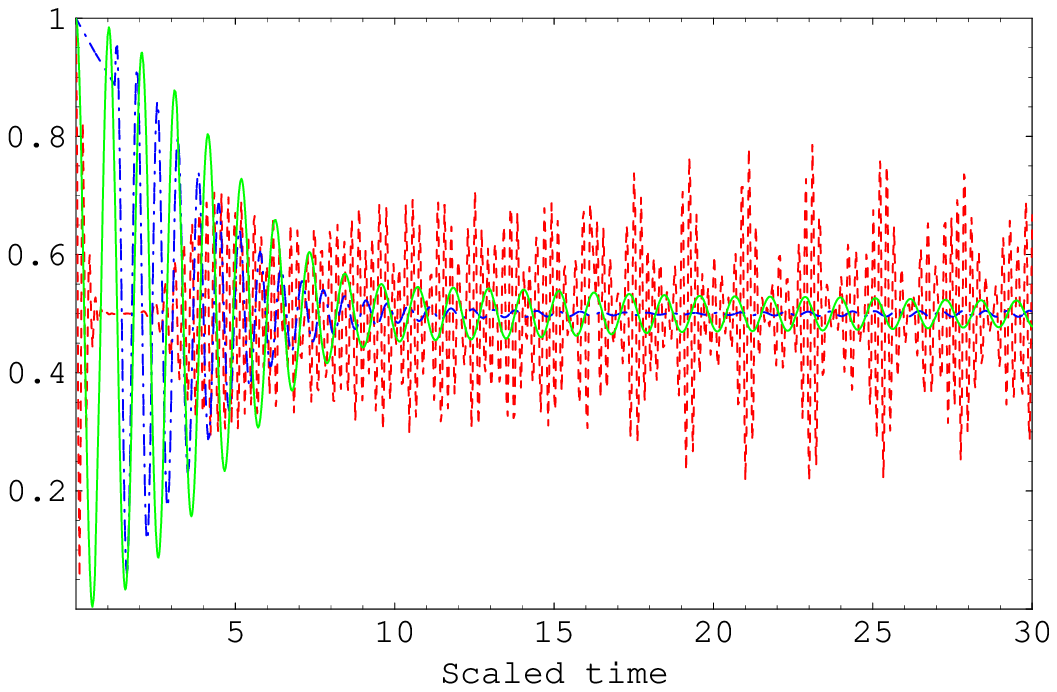}
\put(-220,120){$(c)$} \put(-30,120){$(d)$} \put(-380,75){$\mathcal{F}$%
}\put(-180,75){$\mathcal{F} $}
\end{center}
\caption{The fidelity of the travailing state $|\protect\psi \rangle
$ where $\bar{n}=10$ (a) For the free deformation with $m=1$(b) Free
deformation and
$m=2$ (c)Deformed case where the dash-dot, solid and dot curves for $%
q=0.1,0.5$ and $0.9$ respectively and $m=1$(d)The same as (c) but
$m=2$}
\end{figure}

In the absence of $q$-deformation it is clear that the state $|\phi
\rangle $ lose a large amount of purity compared with that for
$|\psi \rangle $. This is seen in figures (1c) and (2a) where we
also observe that, the purity for
the state $|\psi \rangle $ is slow decaying compared with the second state $%
|\phi \rangle $ which shows fast decaying. In the presences of the
deformation parameter in addition to the existence of more than one
photon inside the cavity, the purity for both states is improved.
Finally for the state $|\psi \rangle $ the minimum and maximum
values are larger than those depicted for the state $|\phi \rangle
$. This means that the state $|\psi \rangle $ is more robust than
the state $|{\phi \rangle }$. Therefore, for the quantum information
tasks it would be much better to find a suitable source to use it
with Bell states of type $|\psi \rangle $ rather than the other Bell
states.

\section{Accuracy of information transfer}

As well known quantum computing depends on transferring information
from one nodes to another one subject to reach the final result.
Therefore, it would be interesting to evaluate the fidelity of the
transmitted information that is carried by the input state. The
fidelity of the output information is given by
\begin{equation}
\mathcal{F}=tr\left( \hat{\rho}_{trans}\hat{\rho}_{out}\right) ,
\label{17}
\end{equation}%
where $\hat{\rho}_{trans}$ refers to the carrier of the transferred
information and $\hat{\rho}_{out}$ is the carrier of the output
information. In figures (3) we exhibit the dynamics of the fidelity
of the transmit information which coded in the traveling state
$|\psi \rangle $. For example, Fig.(3a) displays the behavior of
$\mathcal{F}$ for non-deformed case in presence of a single photon
within cavity, where $m=1$. In this case the fidelity shows behavior
similar to that of the atomic inversion.
\begin{figure}[tbp]
\begin{center}
\includegraphics[width=15pc,height=12pc]{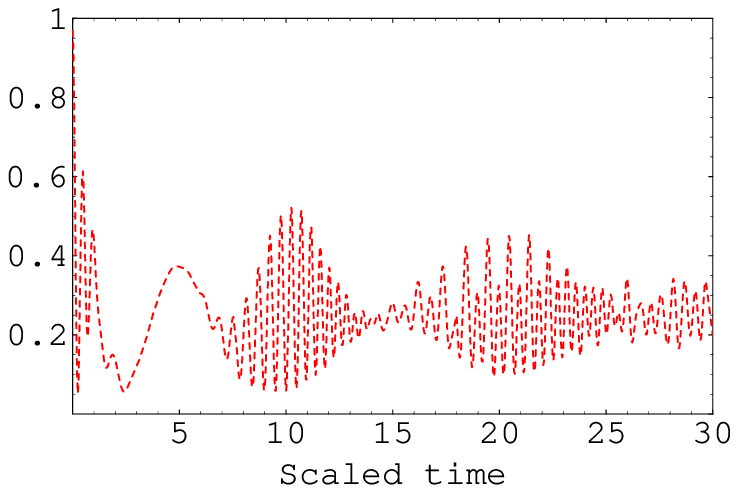} %
\includegraphics[width=15pc,height=12pc]{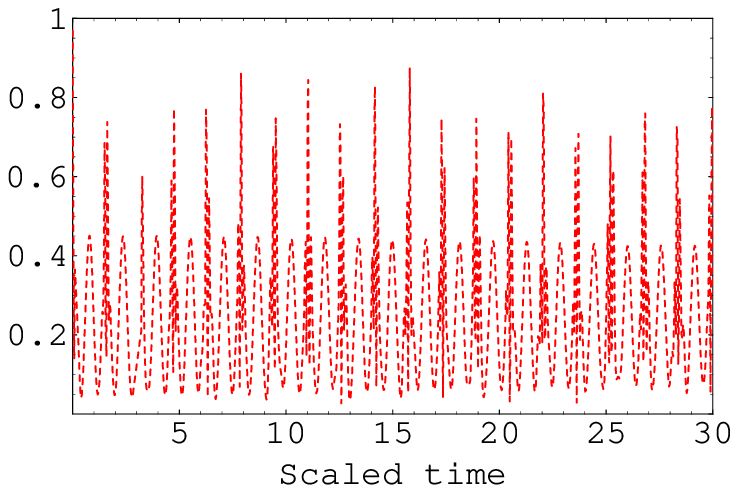}\put(-220,120){$%
(a)$}\put(-30,120){$(b)$} \put(-380,75){$\mathcal{F}$} \put(-180,75){$%
\mathcal{F}$}\\[0pt]
\includegraphics[width=15pc,height=12pc]{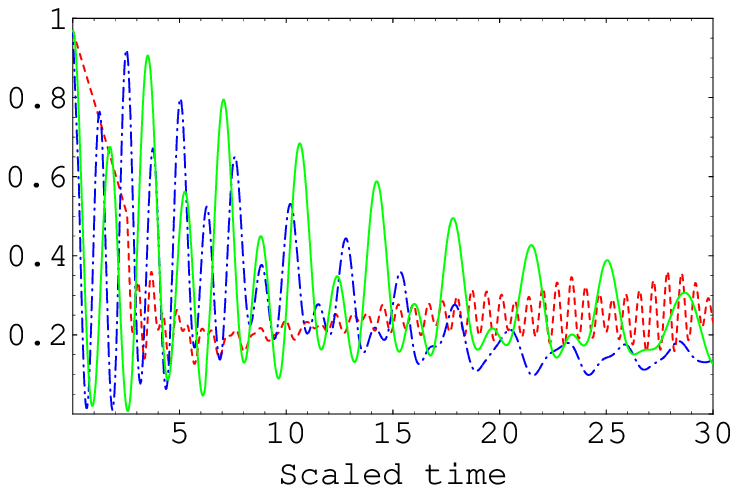} %
\includegraphics[width=15pc,height=12pc]{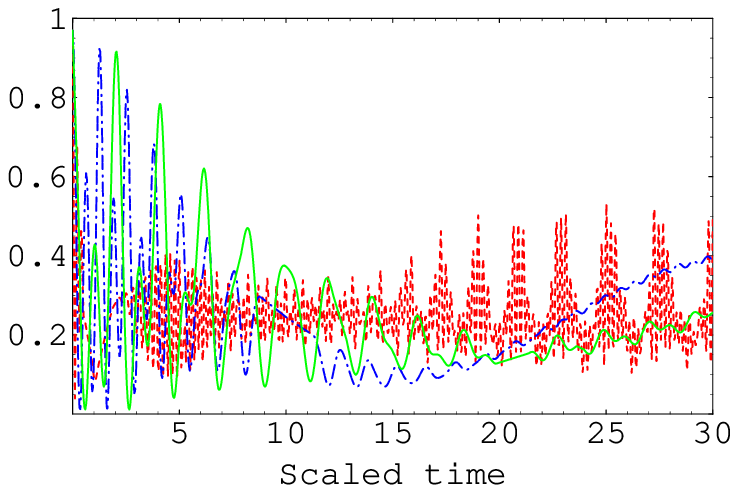}\put(-220,120){%
$(c)$}\put(-30,120){$(d)$} \put(-380,75){$\mathcal{F}$} \put(-180,75){$%
\mathcal{F}$}
\end{center}
\caption{The same as Fig.(3) but for the travailing state $\bigl|\protect%
\phi \bigr\rangle$.}
\end{figure}

This behavior usually appears as a result of the interaction between
the
field and the qubits within cavity. Increasing the number of the photons $%
m=2,$ regular fluctuations can be seen in the fidelity behavior with
an increase in the function amplitude. This means that there is
increasing in the periods of the collapses and revivals as depicted
in Fig.(3b).\

We now turn our attention to consider the case in which $m=1$ to see
the effect of the deformation parameter $q$ on the fidelity
$\mathcal{F}$ of the transfer information. As before, we have
considered three different values of the $q$ parameter $q=0.1,0.5$
and $0.9.$ For the cases in which $q=0.1$ and $q=0.5,$ similar
behavior can be reported, for example the function
fluctuates and decreases its value as the time increases. However, for $%
q=0.1,$ the decreasing is faster than that the case in which
$q=0.5$. Also we have realized that, the amplitude for the case
$q=0.1$ is smaller than
the case $q=0.5$. Moreover, the fluctuations in all the cases including $%
q=0.9$ occur around the value $0.5$. For the case in which $q=0.9$
different behavior can be seen. The function in this case decreases
its value without fluctuations to reach its minimum below the value
$0.4.$ However, after a considerable period of the time compared
with the other two cases. Furthermore, as the time increases the
function shows irregular fluctuations with an increase in its
amplitude, see Fig.(3c). \ When we increase the
value of the photon number and consider $m=2$ we observe that, the function $%
\mathcal{F}$ decreases its value in addition to show rapid
fluctuations after onset of the interaction for the case in which
$q=0.9.$ This is followed with a period of revival and hence the
function turns to display periods of rapid fluctuations with
different amplitudes. In the meantime
different observation can be reported for the other two cases $q=0.1$ and $%
q=0.5$. The function of the fidelity$\mathcal{F}$ in each case shows
an decrease in its value with rapid fluctuations as the time
increases. However, the amplitude of the fluctuations for $q=0.5$ is
greater than that the case for which $q=0.1$, see Fig.(3d).

In Fig.(4), we investigate the dynamics of the fidelity for the
state $|\phi \rangle $. In this case similar behavior can be seen as
shown in Fig.(3).
However, the function reduces its value and fluctuates just above the value $%
0.2$ not $0.5$ as in the previous case. Also there is a small period
of oscillation occurs between two periods of the fluctuations which
is not observed for the state $|\psi \rangle ,$ see Fig.(4a).
Increasing the photon number $m=2,$ more fluctuations can be
reported besides the reduction in function value compared with the
state $|\psi \rangle $ case, see Fig.(4b). Before we close this
section we consider the effect of the $q$-deformation on the
fidelity for the state $|\phi \rangle .$ This is depicted in figures
(4c,d) where the function for the case in which $m=1$ and $q=0.1$
decreases its value faster than that the other two cases $q=0.5$ and
$q=0.9$. Also in
this case the function displays regular oscillations for $q=0.1$ and $q=0.5$%
, while it displays irregular fluctuations for the case in which
$q=0.9$. \
Furthermore, the function decreases its value as the time increases for $%
q=0.1$ and $q=0.5$, but it turns to increase its value slightly for
the case in which $q=0.9$, see Fig.(4c). When we consider the case
$m=2$, we observe for all values of the $q$-deformation parameter
that, there are increasing in the fluctuations compared with the
case in which $m=1$, see Fig.(4d).

In conclusion, the number of photons as well as the deformity play a
role to enhance the fidelity of the transferee information. As it is
shown, the fidelity of the traveling state in the presences of
deformation is better than that the case of the free deformation.
Increasing the number of photons inside the cavity, the fluctuations
of the fidelity increase, however, with a small amplitude. Although
the fidelity decreases as one increases the number of photons in the
presences of the deformation parameter, however it does not tend to
zero as we have shown.

Since the type of the traveling state plays an important role on the
dynamics of the fidelity. Therefore, it is important to compare
different types of input states which carry the required
information. This means that we have to choose which one can resist
during the traveling from one node to another, where the traveling
environment is not perfect.

\section{Populations}

\begin{figure}[b]
\label{Popu1}
\par
\begin{center}
\includegraphics[width=15pc,height=12pc]{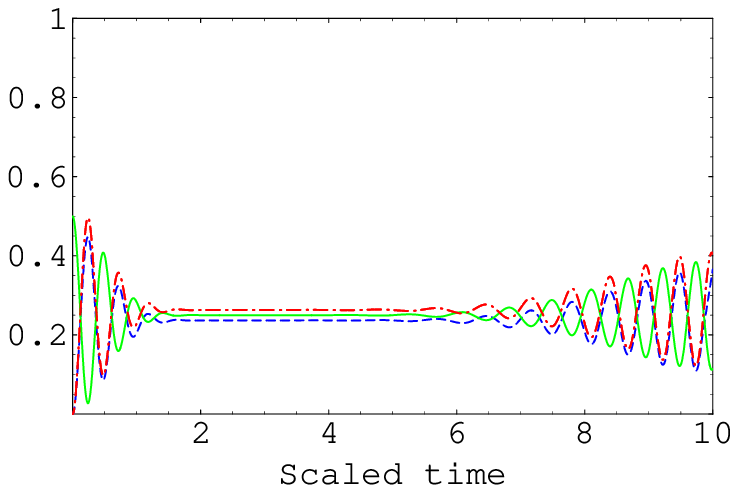} %
\includegraphics[width=15pc,height=12pc]{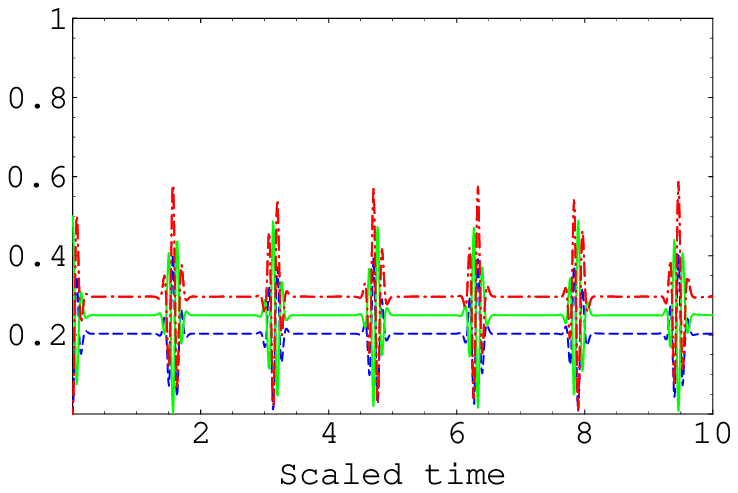} \put(-220,120){$(a)$%
} \put(-30,120){$(b)$} \put(-380,75){$P$} \put(-180,75){$P$}
\end{center}
\caption{The population for the non-deformed case for the travailing state $|%
\protect\psi \rangle $ where $\bar{n}=10$ and $m=1$ (a) The
dash-dot, solid and the dot curves are for $P_{ee},(P_{eg}$ and
$P_{ge})$ and $P_{gg},$ respectively (b) The same as (a) but for
$m=2$.}
\end{figure}

Investigating the separability as well as the entanglement behavior
of the traveling state is important for quantum information
protocols. For this reason we have plotted figures (\ref{Popu1}) for
different values of the
involved parameters that to examine the probability distributions $%
P_{gg},P_{eg},P_{ge}$ and $P_{ee}$ for finding the traveling state
$|\psi
\rangle $ in the states $|g,g\rangle ,|e,g\rangle ,|g,e\rangle $ and $%
|e,e\rangle ,$ respectively. \ In Fig.(\ref{Popu1}a) we display the
dynamics of the populations in absence of the deformation parameter
and in the
presence of a single photon within cavity. It is clear that the populations $%
P_{eg}$ and $P_{ge}$ (solid green curve) are coincide during the
whole period of the interaction, while $P_{ee}$ (dash-dot blue
curve) and $P_{gg}$ (dot red curve) have also the same behavior
during the considered time.
Here we emphasis on the fact that the main difference between $P_{eg}$ , $%
P_{ge}$ and $P_{ee},$ $P_{gg}$ is just phase during the oscillation
periods. This means that the evolvement of the state $|\psi \rangle
$ may takes for a small value of the scaled time the form
\begin{equation}
|\psi \rangle =\alpha \left( |e,g\rangle +|g,e\rangle \right) +\beta
\left( |e,e\rangle +|g,g\rangle \right) .  \label{Str0}
\end{equation}%
However, as time increases the behavior of the populations gets more
stable and the travailed state would takes the form
\begin{equation}
|\psi \rangle =\alpha \left( |e,g\rangle +|g,e\rangle \right) +\beta
_{1}|e,e\rangle +\beta _{2}|g,g\rangle .  \label{Str01}
\end{equation}%
It should be noted that, during the period of the interaction the
travailed
state would switch between the above two states (\ref{Str0}) and (\ref{Str01}%
). In Fig.(\ref{Popu1}b) we display the dynamics of populations for
the case in which $m=2$ in the absence of the $q$-deformation
parameter. In this case the usual phenomena of collapse and revival
are pronounced where the state vector shown stability during the
period of the revival. On the other hand during the period of the
collapse different structure appears where the state $|\psi \rangle
$ completely turns into the state $|\phi \rangle $.

To examine the effect of the $q$-deformation we have plotted figures
(6) for different values of the $q$-deformation parameter as well as
for the photon number $m$. For $m=1$ and $q=0.5$ the functions
$P_{ee}$ and $P_{gg}$ are oscillating with different amplitude,
however, both are in phase. While the functions $P_{eg}$ and
$P_{ge}$ shown similar behavior but they exchange the oscillations
with $P_{ee}$ and $P_{gg}$, see Fig.(6a). When we increase the value
of the photon number, $m=2$ different behavior can be seen. In this
case there are increasing in the number of the oscillations in all
the population functions. However, there are also a clear difference
between each population. For example the population $P_{ee}$
increases its amplitude corresponding to decrease in the amplitude
of the population $P_{gg}.$ Furthermore, as the time increases as
the amplitude of each function decreases. On the other hand we can
observe the same behavior in the function $P_{eg}$, however, with
different phase and amplitude, see Fig.(6b).
\begin{figure}[tbp]
\label{PP2}
\par
\begin{center}
\includegraphics[width=15pc,height=12pc]{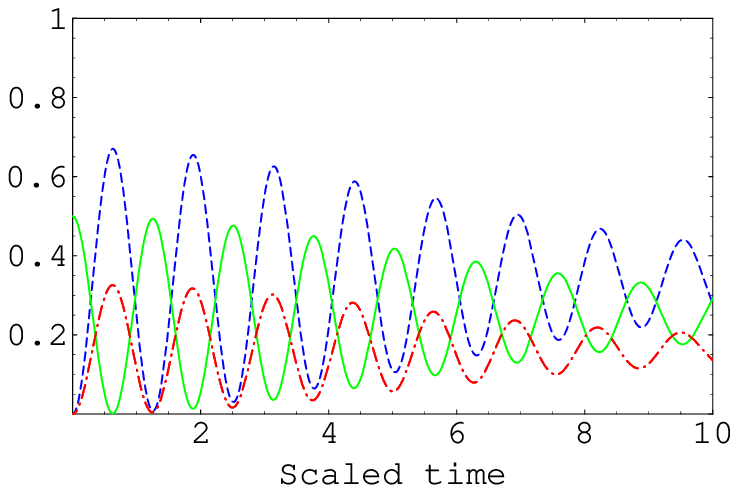} %
\includegraphics[width=15pc,height=12pc]{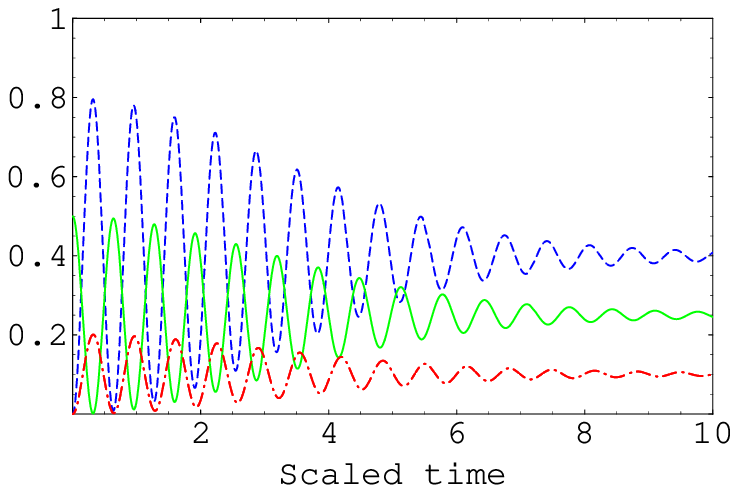} \put(-220,120){%
$(a)$} \put(-30,120){$(b)$} \put(-380,75){$P$} \put(-180,75){$P$}\\[0pt]
\includegraphics[width=15pc,height=12pc]{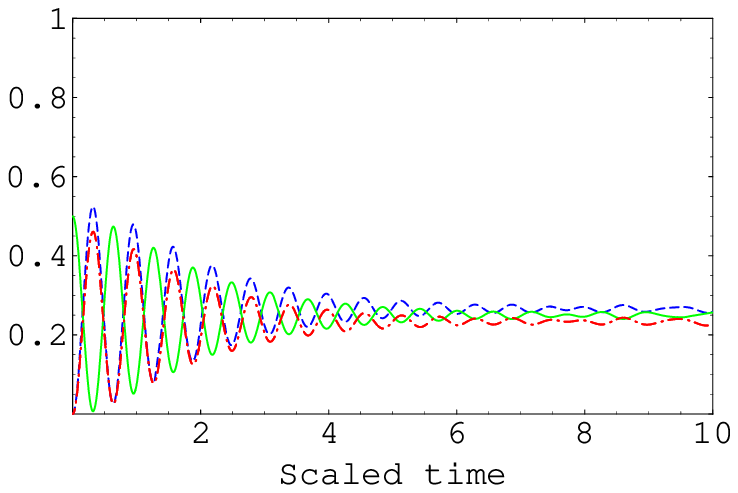} %
\includegraphics[width=15pc,height=12pc]{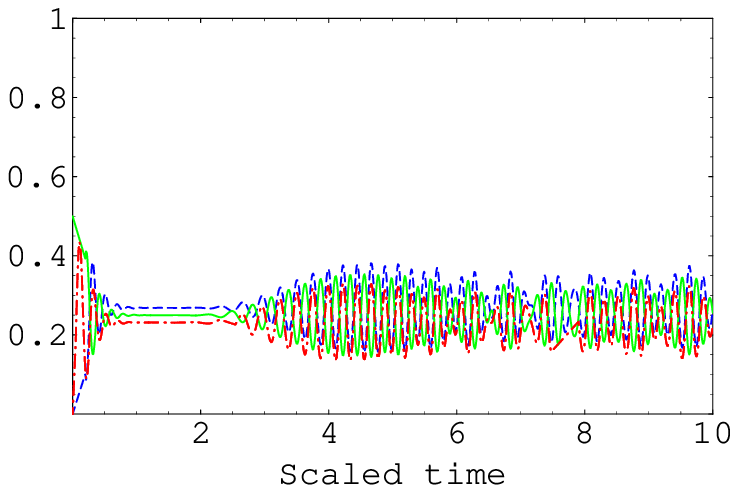} \put(-220,120){%
$(c)$} \put(-30,120){$(d)$} \put(-380,75){$P$} \put(-180,75){$P$}
\end{center}
\caption{The same as Fig.(5), but for the deformed case (a) $m=1$ and $q=0.5$%
(b)$m=2$ and $q=0.5$(c) $m=1$ and $q=0.9$(d)$m=2$ and $q=0.9$}
\end{figure}

When we have examined the case in which $m=1$ and $q=0.9,$ similar
behavior to the case in which $m=2$ and $q=0.5$ is observed.
However, the amplitude for all the population functions are
decreasing as the period of the time increases, see Fig.(6c). In
crease the value of the $q$-deformation, $q=0.9$ leads to rapid
fluctuations in all the population functions. This becomes more
pronounced after certain period of the time, see Fig.(6d). Here we
would like to emphasis on the fact that, as the amplitude of the
populations decreases as the degree of entanglement increases.

Among the important forms which can be detected from figures (6) is
the state
\begin{equation}
|\psi \rangle =\alpha _{3}\left( |e,g\rangle +|g,e\rangle
+|e,e\rangle \right) +\beta _{3}|g,g\rangle ,  \label{Str02}
\end{equation}%
which has a structure similar to that of equation (19) but with the
supper position of the state $|e,e\rangle $ instead of the state
$|g,g\rangle $

\begin{figure}[tbp]
\begin{center}
\includegraphics[width=15pc,height=12pc]{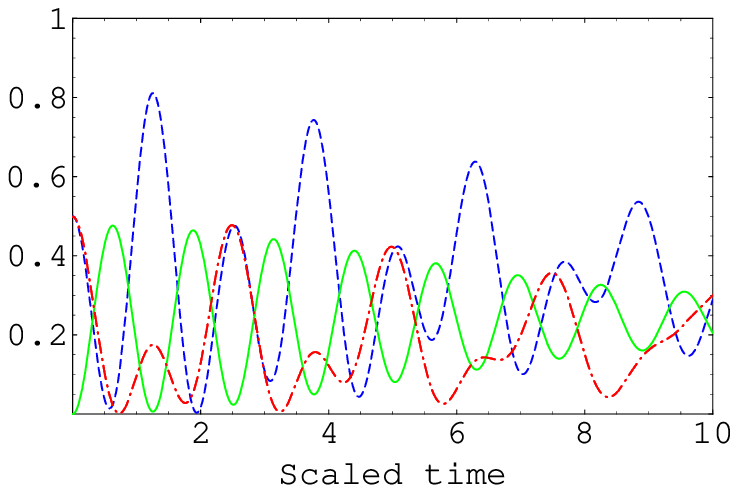} %
\includegraphics[width=15pc,height=12pc]{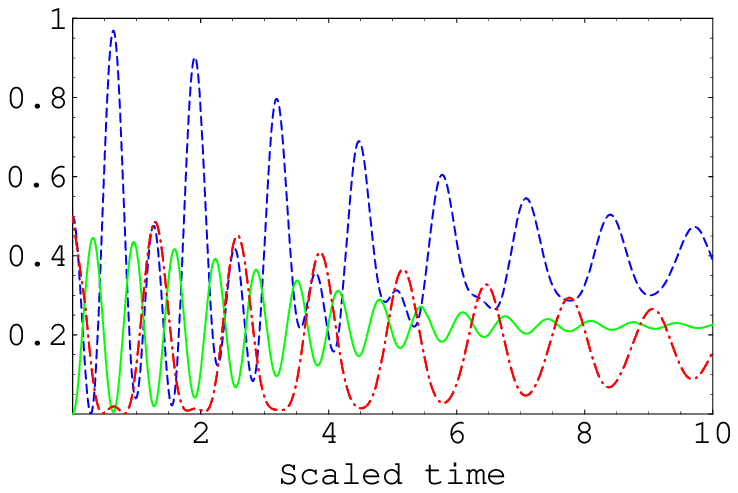} \put(-220,120){$%
(a)$} \put(-30,120){$(b)$} \put(-380,75){$P$} \put(-180,75){$P$}
\end{center}
\caption{The same as Fig.(6a,b) but for the state vector
$\bigl|\protect\phi \bigr\rangle$}
\end{figure}

Finally we investigate the dynamics of the population for the state vector $%
|\phi \rangle $. To do se we have plotted figures (7) for fixed
value of the
$q$-deformation, $q=0.5$ and for $m=1,2$. In this case the populations $%
P_{ee}$ and $P_{gg}$ start with the value 0.5 after onset of the
interaction where both functions show regular oscillations with a
reduction in their values as the time increases. However, for the
first period of the oscillations of the function $P_{gg}$ increases
its value up to $0.8$. On the other hand the function $P_{eg}$ start
with zero value and hence it starts to increase its value up to
$\sim 0.5$ showing regular oscillations. Also we observe that, as
the time increases as the function decreases its value and
consequently the entanglement shows improvement for a large value of
the time, see Fig.(7a). Increasing the value of the photon number,
for example $m=2$ we observe an increase in the value of the
amplitude, more precisely in the function $P_{ee}$ which approaches
the maximum value of the population. In the meantime the function
$P_{gg}$ shows regular fluctuations shrinking as the time increases.
Similar behavior can also reported for the other two populations
$P_{ee}$ and $P_{ge}.$ However, the oscillations in the function
$P_{ge}$ are decaying faster than that the other two populations,
see Fig.(7b).

\section{Conclusion}

In the present paper we have considered the problem of the
interaction between two qubits and multi-photon cavity field. The
creation and annihilation operators which describe the field are
considered to be deformed. An exact analytical solution of the wave
function is obtained and used to construct the density matrix
operator. The effect of the number of photons in addition to
$q$-deformation are examined where the investigation carried out on
the purity, the fidelity of the travailing state and the population.
We have shown that when the number of photons inside the cavity is
increased the fluctuations as well as the amplitude of these
functions are also increased. On the other hand the existence of the
deformation leads to an enhancement of the fidelity and the purity
for small values of the deformity parameter. However for large
values of the $q$-deformation parameter, the minimum values are
always bounded. As a result of the interaction between the qubits
and the field, the structure of the travailing state would takes a
different form. In addition, the nature of the structure can also
causes a decrease in the purity and the fidelity. Therefore, the
evolvement state loses some of its entanglement and turns into
partially entangled state. This in fact encouraged us to examine the
robustness of the travailing state and make a comparison between two
different types of the maximum entangled Bell state. It has been
shown that, the robust state is only pronounced in one of these
states. This means that, coding the information in this robust state
maybe more save during the evolvement to achieve quantum
communication or computation tasks. Since it is possible to control
the devices in the reality, therefore we belief that, the present
results can give us an indication of how the carrier of information
propagate from node to another node.

\textbf{Acknowledgement: }

One of us (M.S.A.) is grateful for the financial support from the
project Math 2010/32 of the Research Centre, College of Science,
King Saud University. 

\end{document}